\documentclass[runningheads,a4paper]{llncs}
\usepackage{graphicx}
\usepackage{amssymb}
\usepackage{wrapfig}
\usepackage{amsmath}
\usepackage{graphicx}
\usepackage[noadjust]{cite}
\usepackage{cite}
\usepackage{url}
\usepackage{epstopdf}
\usepackage{bm}
\usepackage{multirow}
\usepackage[table]{xcolor}
\usepackage[font=small,labelfont=bf,tableposition=top]{caption}
\usepackage{subcaption}
\usepackage[bookmarks=false]{hyperref}
\usepackage[margin=1.2in]{geometry}
\mathchardef\mhyphen="2D
\urldef{\mailsa}\path||
\urldef{\mailsb}\path||
\urldef{\mailsc}\path||    
\newcommand{\keywords}[1]{\par\addvspace\baselineskip
\noindent\keywordname\enspace\ignorespaces#1}

\begin{document}

\title{Human Factors in Homograph Attack Recognition}

\author{Tran Phuong Thao\inst{1}
\and Yukiko Sawaya\inst{2}
\and Hoang-Quoc Nguyen-Son\inst{2}
\and Akira Yamada\inst{2} \and\\
 Ayumu Kubota\inst{2}
\and Tran Van Sang\inst{1}
\and Rie Shigetomi Yamaguchi\inst{1}}
\authorrunning{T. P. Thao et al.}
%
\institute{The University of Tokyo, Japan \\
\email{\{tpthao@yamagula.ic.i, 4040961653@g.ecc, yamaguchi.rie@i\}.u-tokyo.ac.jp}
\and KDDI Research Inc., Japan\\
\email{\{yu-sawaya, ho-nguyen, ai-yamada, kubota\}@kddi-research.jp}}

\titlerunning{Human Factors in Homograph Attack Recognition}

\maketitle

\begin{abstract}
Homograph attack is a way that attackers deceive victims about which website domain name they are communicating with by exploiting the fact that many characters look alike. The attack becomes serious and is raising broad attention when recently many brand domains have been attacked such as Apple Inc., Adobe Inc., Lloyds Bank, etc. We first design a survey of human demographics, brand familiarity, and security backgrounds and apply it to 2,067 participants. We build a regression model to study which factors affect participants' ability in recognizing homograph domains. We find that for different levels of visual similarity, the participants exhibit different abilities. 13.95\% of participants can recognize non-homographs while 16.60\% of participants can recognize homographs whose the visual similarity with the target brand domains is under 99.9\%; but when the similarity increases to 99.9\%, the number of participants who can recognize homographs significantly drops down to only 0.19\%; and for the homographs with 100\% of visual similarity, there is no way for the participants to recognize. We also find that female participants tend to recognize homographs better the male but male participants tend to able to recognize non-homographs better than females. Security knowledge is a significant factor affecting both homographs and non-homographs; surprisingly, people who have strong security knowledge tend to be able to recognize homographs but not non-homographs. Furthermore, people who work or are educated in computer science or computer engineering do not appear as a factor affecting the ability in recognizing homographs; however, interestingly, right after they are explained about the homograph attack, people who work or are educated in computer science or computer engineering are the ones who can capture the situation the most quickly.

\end{abstract}

\section{Introduction}
\label{section:introduction}
Homograph attack is first described by E. Gabrilovic et al.~\cite{homograph_first2002} in 2002. To demonstrate the feasibility of the attack, the authors registered a homograph targeting to the brand domain \url{microsoft.com} using the Russian letters 'c' (U+0421) and 'o' (U+041E). The homograph contains the two non-ASCII characters and has an ASCII converted form as \url{xn--mirsft-yqfbx.com}.\footnote{International Domain Names (IDNs) contain non-ASCII characters (e.g., Arabic, Chinese, Cyrillic alphabet). Therefore, they are encoded to ASCII strings using Punycode transcription known as IDNA encoding and appear under ASCII strings starting with ``\url{xn--}". For example, the domain \url{xn--ggle-0qaa.com} is displayed as \url{g}$\tilde{\text{o}}\tilde{\text{o}}$\url{gle.com}.} However, the attack was not much attracted at that time. Until 2017, the attack had raised broad attention when the famous brand domain \url{apple.com} (Apple Inc.) is attacked by the homograph that appears under the Punycode form~\cite{homograph_apple2017} such as \url{xn--pple-43d.com}, which uses the Cyrillic `a' (U+0430) instead of the ASCII `a' (U+0061). Thereafter, many homograph attacks targeting other famous brand domains have been found such as Adobe Inc.~\cite{adobe}, LLoyds Bank~\cite{Lloydsbank}, Google Analytics~\cite{GoogleAnalytics}, etc. A recent large-scale analysis~\cite{homograph_statistic_dsn2018} about International Domain Names (IDNs) in 2018 shows that, just for the first 1,000 brand domains in top Alexa ranking, more than 1,516 homograph domains were already registered. Furthermore, the attack becomes more progressive and sophisticated today. 

\subsubsection{Motivation}
Many defensive approaches have been proposed such as applying machine learning to some features (e.g., visual similarity metrics, HTML content, and optical character recognition (OCR))~\cite{homograph_visual_screen_shot, homograph_ORC, typosquatting, homograph_IFIP}, using empirical analysis based on registered databases (e.g., Whois, DNS, blacklists, confusable Unicode)~\cite{homograph_statistic_dsn2018, confused_unicode}, or blocking International Domain Names (IDNs) (e.g., disabling the automatic IDN conversion on browsers)~\cite{blocking_1, blocking_2, blocking_3, blocking_4}. So, we ask the question: \emph{how to design an approach that focuses on pro-active defense which can control the attack rather than just responding to it after it has really happened; and is it possible if the approach is based on ergonomics rather than machine engineering?} We therefore in this paper, aim to propose a system that analyzes human factors in the ability of homograph domain identification. This, in turn, allows for various security training courses against the attack aiming to appropriate participants.

\subsubsection{Contribution} 
To the best of our knowledge, our work is the first to devise a system that predicts if human demographics, brand familiarity, and security backgrounds can influence the ability of homograph recognition. To do so, we designed a survey and applied it to 2067 participants who are Internet users in Japan. We subsequently build a regression model to study which factors affect the ability. As a result, we find that for different levels of visual similarity, the participants exhibit different abilities. 13.95\% of participants can recognize non-homographs while 16.60\% of participants can recognize homographs whose visual similarity with the target brand domains is under 99.9\%; but when the similarity increases to 99.9\%, the number of participants who can recognize homographs significantly drops down to only 0.19\%; and for the homographs with 100\% of visual similarity, there is no way for the participants to recognize. We also find that while female participants tend to be able to recognize homographs, male participants tend to able to recognize non-homographs. The result also shows that security knowledge is a significant factor affecting both homographs and non-homographs. We hypothesized that people who have strong security knowledge can recognize both homograph and non-homograph; but surprisingly, it is only true for the case of homographs but not for the case of non-homographs. Another interesting result is that people who work or are educated in computer science or computer engineering do not appear as a factor affecting the ability of homograph recognition. However, right after they are explained about what the homograph attack is, people who work or are educated in computer science or computer engineering are the ones who can capture the situation the most quickly (i.e, from not an affecting factor to become an affecting factor the most quickly). We believe that it opens avenues to help users reduce their presumptuousness and improve knowledge and carefulness about security threats.

\subsubsection{Roadmap}
The rest of this paper is organized as follows. The related work is described in Section~\ref{section:related_work}. The procedure for preparing the survey is presented in Section~\ref{section:procedure}. The methodology is given in Section~\ref{section:methodology}. The experiment is analyzed in Section~\ref{section:experiment}. The discussion is mentioned in Section~\ref{section:discussion}. Finally, the conclusion is drawn in Section~\ref{section:conclusion}.

\section{Related Work}
\label{section:related_work}
In this section, we introduce related work about defending homograph approaches and related work about factor analysis of the brand familiarity, and security background in computer security-related issues.

\subsection{Disabling the Automatic IDN Conversion}
In this approach, the feature of automatic IDN conversion is disabled in the web browser. Instead of showing the converted form of the domain such as \url{g}$\tilde{\text{o}}\tilde{\text{o}}$\url{gle.com}, the browsers only display the original IDN form such as \url{xn--ggle-0qaa.com} in the address bar. In reality, some popular web browsers applied this approach including Chrome and Firefox~\cite{blocking_1}, Safari~\cite{blocking_2}, Internet Explorer~\cite{blocking_3}, and Opera~\cite{blocking_4}. However, there is a big trade-off when the browsers stop supporting the automatic IDN conversion because a large number of Internet users are using non-English languages with non-Latin alphabets through over 7.5 million registered IDNs in all over the world (by December 2017)~\cite{IDN_report2017}. Furthermore, the homograph attack exploits not only look-alike Punycode characters in IDNs, but also look-alike Latin characters in non-IDNs. For instance, the homograph \url{bl0gsp0t.com} targeted to the brand domain \url{blogspot.com} by replacing the ‘o’ by the ‘0’; or the homograph \url{wlklpedia.org} targeted to the brand domain \url{wikipedia.com} by replacing the ‘i’ by the ‘l’. Also, if the homographs can deceive users before appearing in the address bar of browsers (e.g., the homographs are given from an email or a document under hyper-links) without the users' awareness of the browsers, disabling IDN conversion is not meant to prevent users from accessing the homographs. 

\subsection{Detecting Homographs}
Several methods have been proposed to detect homographs. K. Tian et al.~\cite{homograph_visual_screen_shot} scanned five types of squatting domains over DNS records and identified domains that are likely impersonating popular brands. They then build a machine learning classifier to detect homographs using page behaviors, visual analysis and optical character recognition (OCR). L. Baojun et al.~\cite{homograph_statistic_dsn2018} made a large-scale analysis on IDNs using correlating data from auxiliary sources such as Whois, passive DNS and URL blacklist. They found that 1.4 million IDNs were actively registered in which 6000 IDNs were determined as homographs by URL blacklists. They also identified 1,516 IDNs showing high visual similarity to reputable brand domains. S. Yuta et al.~\cite{homograph_ORC} applies machine learning on optical character recognition (OCR) feature of a huge 1.92 million actual registered IDNs and over 10,000 malicious IDNs. A. Pieter et al.~\cite{typosquatting} collected data about the typosquatting homographs of the 500 most popular websites for seven months. They reveal that 95\% of the popular domains they investigated are actively targeted by typosquatters, only few brand owners protect themselves against this practice by proactively registering their own typosquatting domains. The study also reveals that a large  of typosquatting homographs can be traced back to a small group of typosquatting page hosters and that certain top-level domains are much more prone to typosquatting than others. T. Thao et al.~\cite{homograph_IFIP} constructed a classification model for homographs and potential homographs registered by attackers using machine learning on feasible and novel features which are the visual similarity on each character and selected information from Whois. Several tools~\cite{dnstwist,idn_homograph_attack,EvilURL,homographs_dutch,instant_domain_search,DN_Pedia,Homoglyph_Attack} generate permutations of homographs from a defined subset of look-alike characters from Confusable Unicode table defined by Unicode Inc.~\cite{confused_unicode}, then look up Whois and DNS to check whether the homographs are registered and active. Compared to the approach of disabling the automatic IDN conversion, the homograph detection is more attractive to the research community. 

\subsection{Brand Familiarity and Security Backgrounds in Computer Security}
In this section, we present work related to web familiarity and security backgrounds including security warnings, security knowledge, security behavior, and security self-confidence that affect human decisions on security threats. Since some previous papers analyzed both brand familiarity and security backgrounds, we do not separate them into two different sections. 

T. Kelley et al.~\cite{nihgov} simulate several secure non-spoof and insecure spoof domains with different authentication levels such as extended validation, standard validation, or partial encryption. A logistic model is then applied to participants' respondents to compare how encryption level, web familiarity, security knowledge, and mouse tracking influence the participant accuracy in identifying spoof and non-spoof websites. Their result shows that user behavior derived from mouse tracking recordings leads to higher accuracy in identifying spoof and non-spoof websites than the other factors. Y. Sawaya et al.~\cite{chi2017} apply the Security Behavior Intentions Scale (SeBIS)~\cite{SeBIS} to participants from seven countries and build a regression model to study which factors affect participants' security behavior using a cross-cultural survey. The work concluded that self-confidence in computer security has a larger positive effect on security behavior compared to actual knowledge about computer security. I. Kirlappos et al.~\cite{phishing_sp2012} show that users do not focus on security warnings (or not understand what they are) rather than looking for signs to confirm whether a site is trustworthy. The study reveals that advice given in some current user educations about phishing is largely ignored. It, therefore, suggests that rather than flooding users with information, we need to consider how users make decisions both in business and personal settings for the user education. M. Sharif et al.~\cite{urakawa_ccs2018} design a survey of security warnings, user behavior, knowledge and self-confidence about security to evaluate the utility of self-reported questionnaire for predicting exposure to malicious content. Their result confirms that the self-reported data can help forecast exposure risk over long periods of time but is not as crucial as behavioral measurements to accurately predict exposure. S. Das et al.~\cite{soups_social} find that social processes played a major role in security behavior. Furthermore, conversations about security are often driven by the desire to warn or protect others from immediate novel threats observed or experienced. C. Erika et al.~\cite{soups_mobile} study user confidence toward security and privacy for smartphone and find that participants are apprehensive about running privacy- and financially-sensitive tasks on their phones as four factors: fear of theft and data loss, misconceptions about the security of their network communications, worries about accidentally touching or clicking, and mistrust of smartphone applications. I. Iulia et al.~\cite{soups_expert} compare security behaviors between expert and non-expert and find that while experts frequently report installing software updates, using two-factor authentication and using a password manager, non-experts report using antivirus software, visiting only known websites, and changing passwords frequently. A. Felt et al.~\cite{android} examine whether security warnings from the Android permission system is effective to users. Their result shows that only 17\% of participants paid attention to permissions during installation, and only 3\% of Internet survey respondents could correctly answer all permission comprehension questions. This indicates that current Android security warnings do not help most users make correct security decisions.

\section{Procedure}
\label{section:procedure}
In this section, we present how the survey is designed and distributed to the participants. The survey is created in the Japanese and is embedded to a webpage. The webpage is then distributed to 2,067 participants who are Internet users in Japan.\footnote{The Appendix in this paper describes the questions in English but the survey is designed in Japanese language and distributed to Japanese, so there is no translation problem for the preservation of the survey's reliability and structure validity.} The participants cannot submit their responses if any of the questions is not answered.  There are three question parts about the human factors (including demographics, brand familiarity, and security backgrounds), and the final part about the participants' ability in distinguishing homographs. The following sections describe the design of each part.

\subsection{Demographics}
For the human demographics, the survey consists of the following seven questions:
\begin{enumerate}
\item Gender (male: 1 and female: 0)
\item Age (the inputs are integers)
\item Having a job (having a full-time job: 1, freelancer or part-time job: 0.5, and not have a job: 0).
\item Whether the participant has studied so far the languages including English, Spanish, French, Russian, Portuguese, German, Vietnamese, Turkish, Italian, Greek, and Dutch. The languages chosen are the common languages that use Punycode (i.e., confusable letters with the English alphabet). For each language, there are two answer options (yes:1 and no: 0). Thereafter, we calculate the number of languages that the participants answer `yes'. 
\item Knowing only Japanese (yes: 1, and no: 0). Although there is a variable related to the number of languages that the participants have studied so far, we hypothezied that knowing only Japanese or not is probably an affecting factor because the survey is done in Japan. Thereby knowing only Japanese is chosen as a variable that needs to be measured.
\item Whether the participant graduated or enrolled in computer science or computer engineering (yes: 1 and no: 0).
\item Whether the participant worked (or is working) in computer science or computer engineering (yes: 1 and no: 0).
\end{enumerate}

\subsection{Brand Familiarity}
For the brand familiarity, the nine famous brands are chosen including Amazon, Google, Coinbase, Wiki, Booking, Expedia, Paypal, Sex.com and Facebook. For each of the brands, the participants respond to how they are familiar with the brands with 4-point Likert-scale answer options (do not know: 1, know but never use: 2, occasionally use: 3, and often use: 4). The brands may have multiple authentic domains (i.e., the domains that the brands themselves registered), and thus the logos and the names of the brands are used to represent the brands and showed in the questions instead of listing all their domains.  

\subsection{Security Backgrounds}
For the security backgrounds, the survey consists of the following five questions:
\begin{enumerate}
\item Anti-virus software installation on PCs or mobile devices: (yes: 1 and no: 0).
\item Security warning: When browsing a website, a browser or anti-virus software issues a warning, whether the participants continue browsing or not (yes: 1 and no: 0).
\item Security behavior: that consists of sixteen sub-questions as described in Appendix~\ref{section:security_behavior}. For each of the sub-questions, the participants choose 5-point Likert-scale answer options (not at all: 1, rarely: 2, sometimes: 3, often: 4, and always: 5). The summation of all the sixteen answers is then calculated and used as the variable in the model instead of each separated answer.
\item Security knowledge: that consists of eighteen sub-questions as described in Appendix~\ref{section:security_knowledge}. For each of the sub-questions, the participants have two answer options (true: 1 and false: 0). Then, based on the actual correct answers given at the end of the appendix, we count the number of correct answers of the participants. 
\item Security self-confidence: that consists of six sub-questions as described in Appendix~\ref{section:security_confidence}. The participants have 5-point Likert-scale answer options (not at all: 1, not applicable: 2, neither agree nor disagree: 3, applicable: 4, and very applicable: 5). Similar to the security behavior, the summation of the six answers is calculated and used for the model.
\end{enumerate}
For the security behaviors, security knowledge, and security self-confidence, we use the design from the paper~\cite{chi2017}. The paper aims to analyze factors that affect security behavior and thus uses security behavior in the target function. Meanwhile, our work aims to analyze factors (including security behavior) that affect the ability of homograph recognition, and thus security behavior is just one of the features, not used in the target function. 

\begin{figure}[!htb]
\center
\includegraphics[width=0.7\columnwidth]{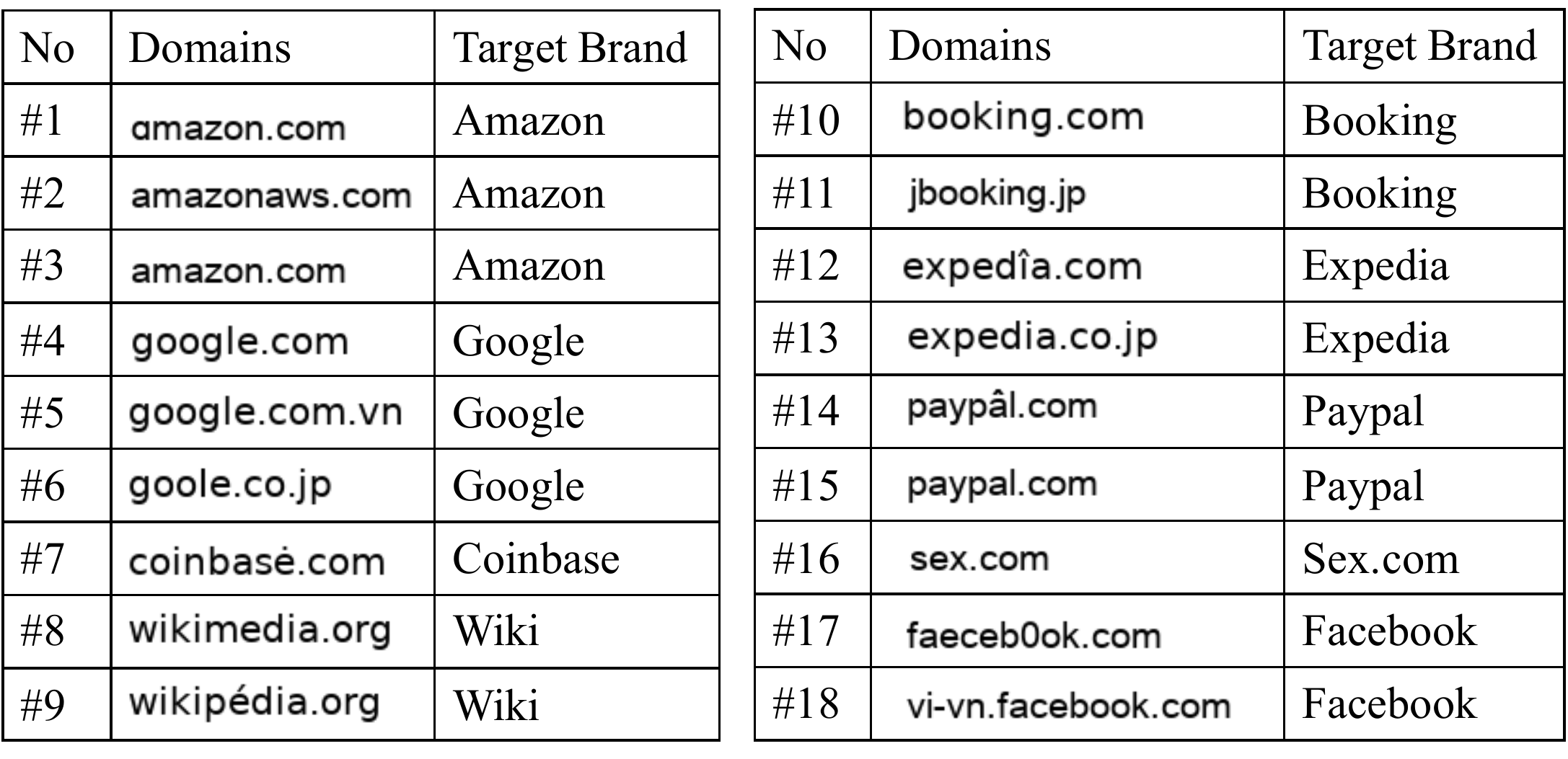}
\caption{Sample Domains Used for Testing the Ability in Distinguishing Homographs}
\label{fig:user_decision} 
\end{figure}

\subsection{Homograph Recognition}
This part is used for calculating the values of the target function. The eighteen sample domains mixed between homographs and non-homographs are showed in Figure~\ref{fig:user_decision} and explained in Appendix~\ref{section:user_decision}. The domains target to the nine brands mentioned in the brand familiarity. The domains are chosen for different purposes. For example, the domain \#2 (\url{amazonaws.com}) is chosen because participants probably only know \url{amazon.com} and think \url{amazonaws.com} is a homograph but actually it is not. Another example is the domain \#16 (\url{sex.com}) which is a pornographic domain, and thus the participants probably think it is homograph (unsafe) but actually it is not. For each of the eighteen domains, the participants answer whether it is safe or not. Based on the correct answers described in the Appendix~\ref{section:user_decision}, we extract whether the participants have a correct answer for each domain (true: 1, and false: 0). The reason we choose the number of domains as 18 but not 30, 40 or even more is that the participants will tend to randomly choose the answer options instead of actually answering if a survey contains too many questions, and 18 questions are a good limit for our design. 

\section{Methodology}
\label{section:methodology}
This section describes the pre-process on the raw data of the participants' responses, determine the target function and define the model.

\subsection{Domain Grouping}
\label{section:grouping}
The eighteen sample domains are grouped based on the visual similarity with the brand domains. In this paper, the Structural Similarity Index (SSIM)~\cite{ssim2004} is chosen for the visual similarity metric. SSIM is commonly used since it outperforms the traditional methods such as Peak Signal-To-Noise Ratio (PSNR) and Mean Squared Error (MSE) which can estimate only the absolute errors. Firstly, the domains are parsed to images in the same size  $N \times N$. The SSIM between two images $x$ and $y$ is then calculated as follows:

\begin{equation}
SSIM(x,y) = \frac{(2\mu_x\mu_y+c_1)(2\sigma_{xy}+c_2)}{(\mu^2_x+\mu^2_y+c_1)(\sigma^2_x+\sigma^2_y + c_2)}
\end{equation}
The $\mu_x$ and $\mu_y$ represent the averages of $x$ and $y$, respectively. The $\sigma^2_x$ and $\sigma^2_y$ represent the variances of $x$ and $y$, respectively. $c_1 = (k_1L)^2$ and $c_2 = (k_2L)^2$ represent the variables to stabilize the division with weak denominator where $L$ is the dynamic range of the pixel-values and is typically set to $L = 2^{\#bits\_per\_pixel}-1$ and $k_1 = 0.01, k_2 = 0.03$ by default. SSIM values $[-1, 1]$ where 1 indicates perfect similarity. 

\begin{table}[!h]
\center
\caption{The SSIM of Eighteen Sample Domains}
\begin{tabular}{|>{
\hspace{0.2pc}}c<{\hspace{0.2pc}}
|>{\hspace{0.2pc}}c<{\hspace{0.2pc}}
|>{\hspace{0.2pc}}c<{\hspace{0.2pc}}
|>{\hspace{0.2pc}}l<{\hspace{0.2pc}}
|>{\hspace{0.2pc}}r<{\hspace{0.2pc}}
|}
\hline
\textbf{Group No} & \textbf{Group Name}& \textbf{Domain\#} & \textbf{Brand Domain} & \textbf{SSIM}\\
\hline
\hline
\multirow{ 4}{*}{Group 1}  & \multirow{ 4}{*}{\shortstack{Homographs with \\SSIM $\geq 0.999$}}  & {\#3\ } & \url{amazon.com} & 1.000\\
        & & {\#4\ } & \url{google.com} & 1.000\\
        && \#10 & \url{booking.com} & 0.999\\
        && \#15 & \url{paypal.com} & 1.000\\
\hline
\multirow{ 7}{*}{Group 2} & \multirow{ 7}{*}{\shortstack{Homographs with \\SSIM $< 0.999$}} & {\#1\ } & \url{amazon.com} & 0.994\\
        && {\#6\ } & \url{google.com} & 0.838\\
        && {\#7\ } & \url{coinbase.com} & 0.996\\
        && {\#9\ } & \url{wikipedia.org} & 0.994\\
        && \#12 & \url{expedia.com} & 0.995\\
        && \#14 & \url{paypal.com} & 0.993\\
        && \#17 & \url{facebook.com} & 0.845\\
\hline
\multirow{ 7}{*}{Group 3}& \multirow{ 7}{*}{Non-homographs}  & {\#2\ } & \url{amazon.com} & 0.865\\
        && {\#5\ } & \url{google.com} & 0.950\\
        && {\#8\ } & \url{wikipedia.org} & 0.853\\
        && \#11 & \url{booking.com} & 0.780\\
         && \#13 & \url{expedia.com} & 0.950\\
        && \#16 & \url{sex.com} & 1.000\\
        && \#18 & \url{facebook.com} & 0.667\\
\hline
\end{tabular}
\label{table:SSIM_domain}
\end{table}

Using the SSIM, the eighteen sample domains are categorized into three groups:

\begin{itemize}
\item \emph{Group 1: Homographs with SSIM $\geq$ 0.999.}
This group consists of four homographs including the domains \#3, \#4, \#10 and \#15 in Figure~\ref{fig:user_decision}. The domains \#3, \#4 and \#15 have SSIM = 1 which means they look completely the same as the brand domains. The domain \#10 has SSIM = 0.999 because the look-alike letter `g' is very difficult to be recognized.
 
\item \emph{Group 2: Homographs with SSIM $<$ 0.999.}
This group consists of seven homographs including domains \#1, \#6, \#7, \#9, \#12, \#14, and \#17 in Figure~\ref{fig:user_decision}. This group considers the homographs whose SSIM scores are lower than those in Group 1, but not so low, i.e., ranging from 0.838 to 0.996. Other homographs with lower SSIM are not considered since it may be trivial to be recognized by the participants.

\item \emph{Group 3: Non-homographs.}
This group consists of seven non-homographs including the domains \#2, \#5, \#8, \#11, \#13, \#16, and \#18 in Figure~\ref{fig:user_decision}. The domains \#2, \#5, \#8, \#11, \#13, and \#18 are safe domains that are registered by the brand themselves for different services but have less popularity than the main brand domains. For instance, the domain \#2 \url{amazonaws.com} (Amazon Web Services (AWS)) is a cloud computing service of Amazon. Many people may be confused with the main service of Amazon which is the selling service \url{amazon.com}. The domain \#16 \url{sex.com} is chosen since we want to know how participants balance their decisions between a domain that is famous and actually safe with a domain that is notorious for its content category (e.g., pornographic, darknet, terrorism). 
\end{itemize}
For each group, the domain numbers, the brand domains, and the corresponding SSIMs are summarized in Table~\ref{table:SSIM_domain}.

\subsection{Lucky Answers and Neutral Answers}
\label{section:luckyneutral}
The survey is designed so that for each of the eighteen sample domains, the participants not only answer whether the domain is a homograph but also describe the reasons for their decision. A lucky answer is an answer that has a correct decision but inappropriate reason. A neutral answer is an answer that has a correct decision but unclear reason. For instance, a participant who decides \url{goole.co.jp} as a homograph and answers a correct reason such as ``\emph{the letter g is missing}" is not considered as a lucky answer. A participant who decides \url{goole.co.jp} as a homograph and answers an incorrect reason such as `\emph{Google only has .co.jp as a top-level domain, and thus google.com is unsafe}" is considered as a lucky answer. A participant who decides \url{goole.co.jp} as a homograph and answers an unclear reason such as ``\emph{I have a feeling that}" is considered as a neutral answer.

\begin{table*}
\centering
\caption{Lucky Answers and Neutral Answers}
\begin{tabular}{
|>{\hspace{0.2pc}}c<{\hspace{0.2pc}}
|>{\hspace{0.2pc}}c<{\hspace{0.2pc}}
|>{\hspace{0.2pc}}r<{\hspace{0.2pc}}
|>{\hspace{0.2pc}}r<{\hspace{0.2pc}}
|>{\hspace{0.2pc}}r<{\hspace{0.2pc}}
|}
\hline
\multirow{ 2}{*}{\textbf{Group}} & \multirow{2}{*}{\textbf{Domain \#}}  & \multirow{ 2}{*}{\textbf{Incorrect Answer}}  & \multicolumn{2}{c|}{\textbf{Correct Answer}}\\
\cline{4-5}
& && \textbf{Appropriate Reason} & \textbf{Lucky and Neutral Answers}\\
\hline
\hline
\multirow{ 4}{*}{Group 1} & {\#3\ } & 1411 (68.26 \%) & 0 (\hphantom{0}0\hphantom{.00} \%) & 656 (31.74 \%)\\
& {\#4\ } & 1432 (69.28 \%) & 0 (\hphantom{0}0\hphantom{.00} \%)& 635 (30.72 \%)\\
& \#10 & 755 (36.53 \%) & 4 (\hphantom{0}0.19 \%)& 1308 (63.28 \%)\\
& \#15 & 756 (36.57 \%)& 0 (\hphantom{0}0\hphantom{.00} \%) & 1311 (63.43 \%)\\
\hline
\multirow{ 7}{*}{Group 2} & {\#1\ } & 495 (23.95 \%) &  470 (22.74 \%)& 1102 (53.31 \%) \\
& {\#6\ } & 649 (31.40 \%) & 167 (\hphantom{0}8.08 \%)& 1251 (60.52 \%) \\
& {\#7\ } & 173 ( 08.37 \%)& 302 (14.61 \%)& 1592 (77.01 \%) \\
& {\#9\ } & 354 (17.13 \%) & 296 (14.32 \%) & 1417 (68.55 \%) \\
& \#12 & 243 (11.76 \%)& 341 (16.50 \%) & 1483 (71.75 \%)\\
& \#14 & 171 ( 08.27 \%)& 354 (17.13 \%)& 1542 (74.60 \%) \\
& \#17 & 229 (11.08 \%) & 471 (22.79 \%)& 1367 (66.13 \%)\\
\hline
\multirow{ 7}{*}{Group 3}  & {\#2\ } & 1796 (86.89 \%)&   \multicolumn{2}{c|}{271 (13.11 \%)} \\
& {\#5\ } & 1823 (88.20 \%) & \multicolumn{2}{c|}{244 (11.80 \%)}  \\
& {\#8\ } & 1827 (88.39 \%) & \multicolumn{2}{c|}{240 (11.61 \%)} \\
& \#11 & 1832 (88.63 \%)& \multicolumn{2}{c|}{235 (11.37 \%)} \\
& \#13 & 1397 (67.59 \%)& \multicolumn{2}{c|}{670 (32.41 \%)} \\
& \#16 & 1841 (89.07 \%)& \multicolumn{2}{c|}{226 (10.93 \%)} \\
& \#18 & 1935 (93.61 \%)& \multicolumn{2}{c|}{132 (\hphantom{0}6.39 \%)}\\
\hline
\end{tabular}
\label{table:lucky}
\end{table*}

The lucky answers are excluded from the dataset since they are completely data outliers. For the neutral answers, we cannot just flip the decision from \emph{true} to \emph{false} because there is a well-known finding from researchers showing that very often, human experts cannot explain why they make a choice that they do, but they are correct far more often than non-experts. Ericsson et al.~\cite{lucky1} first found this studying chess experts, and the finding has been replicated and found many times since then by people such as Gerd Gigerenzer et al.~\cite{lucky2} and Gary Klein~\cite{lucky3}. This means that it is difficult to classify the neutral answers into data bias or actual correct answers. Therefore, in this paper, we decide to just exclude them from the dataset. It is safe rather than adjusting the participant responses like flipping from \emph{true} to \emph{false}. We manually check each of $2,067\times18$ answers from the 2,067 participants for the eighteen sample domains to find the lucky answers and neutral answers and summarize in Table~\ref{table:lucky}. In this table, the incorrect answers (column 3) and the correct answers with appropriate reasons (column 4) are used for the model. For group 3 (non-homograph), we do not need to remove lucky and neutral answers because: if the participants answer correctly (i.e., the domains are non-homograph), there is nothing to do; but if they answer incorrectly (the domains are homograph), with any reason, the participant's decisions are wrong.  

\subsection{Model}
Let $f$ denote the model for measuring the participants' ability in distinguishing homographs. $f$ is defined as follows:

\begin{equation}
f\sim \mathsf{Demographics} + \mathsf{WebFamiliarity} + \mathsf{SecBackgrounds}
\end{equation}
The explanatory variables related to $\mathsf{Demographics}$ consist of gender, age, having a job, whether the participants know only Japanese, the number of specific languages that the participant has studied so far, whether the participant is educated in computer science/ computer engineering, whether the participant works in computer science or computer engineering. The explanatory variable related to $\mathsf{WebFamiliarity}$ is the usage frequency of the brands. The explanatory variables related to $\mathsf{SecBackgrounds}$ are anti-virus installation, security warnings, security behaviors, security knowledge, and security self-confidence. 

\subsubsection{Target Functions}
The incorrect answers and the correct answers with appropriate reasons are extracted for the model. For each group, two experiment plans are performed using two different target functions. 

\begin{itemize}

\item \emph{Integration}:
This plan integrates all the domains in the group using the target function:
\begin{equation}
\label{eq:target1}
f_1 = \sum_{d_i} \mathsf{SSIM}(d_i, b_i) \times \mathsf{difficult}(d_i) * \mathsf{decision}(d_i)
\end{equation}
where $\mathsf{decision}(d_i)$ denotes the decision of the participants in distinguishing whether the domain $d_i$ is a homograph. $\mathsf{SSIM}(d_i, b_i)$ denotes the SSIM between the domain $d_i$ and its corresponding brand domain $b_i$. $\mathsf{difficulty}(d_i)$ denotes the difficulty of the domain $d_i$ and is defined as ($1-\frac{c_i}{t}$) in which $c_i$ is the number of participants who give correct decisions for $d_i$ and $t = 2,067$ is the total number of participants. For example, there are 10 participants in which 7 participants answer correctly and thus the difficulty of the question is $1-\frac{7}{10}$. In this plan, the multiple (linear) regression model is applied one time for all the domains, and then the affecting factors for the integration target functions are extracted. 

\item \emph{Separation}:
This plan applies the multiple (linear) regression model for each domain in the group and finds the affecting factors for each domain. The common affecting factors are then extracted. The target function is defined as follows:
\begin{equation}
\label{eq:target2}
f_2 = \mathsf{decision}(d_i)
\end{equation}
Since each domain is considered separately, $\mathsf{SSIM}(d_i, b_i)$ and $\mathsf{difficult}(d_i) $ are not necessary for the target function. After the factors affecting the target function are determined, the common factors for all the domains are extracted. 
\end{itemize}

The SSIM and the difficulty are not used as variables in the features but used as elements in the target functions because the SSIM and the difficulty are not related to human information but domain information, and the goal in this paper is analyzing the human factors. Furthermore, for each domain, the SSIM and the difficulty are the same for all 2,067 participants. If the SSIM and the difficulty are used as the variables, the regression model always results that the SSIM and the difficulty are the affecting factors with $p \leq 0.05$. It is therefore not meant in finding factors.

\subsubsection{Factor Determination}
Before showing how the factors affecting the target functions are determined, we briefly describe the preliminary of the (student) $t$-test. A $t$-test~\cite{t-test1, t-test2} is commonly used to determine whether the mean of a population significantly differs from a specific value (called the hypothesized mean) or from the mean of another population. In other words, the $t$-test can tell if the differences could happen by chance. For the first step, the $t$-test takes the sample from each set and establishes the problem statement by assuming a null hypothesis that the two means are equal. Then, it calculates certain values and compares them with the standard values to determine if the assumed null hypothesis is accepted or rejected. If the null hypothesis is rejected, it indicates that data readings are strong and are not by chance. In the $t$-test, the $t$-value represents a ratio between the difference between the two groups and the difference within the groups. The larger the $t$-value, the more difference there is between groups (the more likely it is that the results are repeatable). The smaller the $t$-value, the more similarity there is between groups. If the $t$-value is negative, it shows a reversal in the directionality of the effect being studied. However, it has no impact on the significance of the difference between groups of data. Every $t$-value has a corresponding $p$-value. A $p$-value is the probability that the results from the sample data occurred by chance. The $p$-values vary from 0 to 1. The low $p$-value is good (it indicates the data did not occur by chance). In most cases, a $p$-value that is $\leq 0.05$ is accepted to mean the data is valid. In this paper, the affecting factors have the following $p$-values: 
\begin{itemize}
\item $p \leq 0.001$: \emph{significant factors} that strongly affect the target function, marked as (***) in the experiment result.
\item $0.001 < p \leq 0.01$: \emph{semi-significant}, marked as (**) in the experiment result.
\item $0.01 < p \leq 0.05$: \emph{normal factor} affecting the target function, marked as (*) in the experiment result.
\end{itemize}
In the experiment result, we also show 95\% confidence interval (CI) which is a range of likely of the unknown population parameter. For the first plan (integration), the common samples which contain only incorrect answers and correct answers with appropriate reasons are inputted in the regression model. The factors are then determined based on the $t$-test's result. For the second plan (separation), the factors affecting the target function in each domain are determined. The common factors are then extracted. The final factors chosen for this plan is the common factors that affect $\geq \lceil \frac{N}{2} \rceil$ domains where $N$ denotes the number of domains in the group, and $\lceil \frac{N}{2}  \rceil$ denotes the upper bound of $\frac{N}{2} $. 

\subsection{Consistency of Integration and Separation Plans}
\label{section:consistency}
The best case is when both the plans result in the same set of affecting factors. If the case does not happen, we determine the final affecting factor as follows. Let $I$ and $S$ denote the set of affecting factors found in the integration and separation plan, respectively. Let $R$ denote the set of the affecting factors that we are aiming to find.
\begin{itemize}
\item All the common affecting factors of both the plans $I \cap S$ are included in $R$.
\item If there exists a factor $x \in I$ such that $x \notin S$, $x$ is included in $R$ if $x$ is an \emph{significant factor} in the integration plan ($p \leq 0.001$). 
\item If there exists a factor $x \in S$ such that $x \notin I$, $x$ is included in $R$ if the \emph{significant} $p$-values ($p\leq 0.001$) are dominant in $S$ (i.e., the significant $p$-values belong to more than $\frac{\mid S \mid}{2}$ domains where $\mid S \mid$ denotes the number of factors in $S$).
\end{itemize}
The consistency of both the plans is the final result used for the conclusion; however, the factors found in each plan still gives a lot of important information and we cannot omit their details. 

\section{Experiment}
\label{section:experiment}
The program is written in Python 3.7.4 on a computer MacBook Pro 2.8 GHz Intel Core i7, RAM 16 GB. The multiple (linear) regression model is executed using \emph{scikit-learn} package version 0.21. The $t$-test is computed using \emph{statsmodels} package version 0.10. The SSIM is computed using the \emph{skimage} package version 0.15.dev0.

\subsection{Participant Population}
Before performing the model, we check if the participant sampling process is valid. First, we analyze whether the participant demographics of gender and age statistically match those of an actual data (e.g., data from government census). Second, we show that the distribution of the age (continuous values) is a normal distribution (Gaussian distribution) (for the gender, the data is binary not continuous values and thus, there is no need for normal distribution test).

\begin{table*}
\centering
\captionsetup[subtable]{position = below}
\captionsetup[table]{position=top}
\caption{Participant Sampling}
\label{table:population}
\begin{subtable}{0.3\linewidth}
\centering
\begin{tabular}{
|>{\hspace{0.2pc}}l<{\hspace{0.2pc}}
|>{\hspace{0.2pc}}r<{\hspace{0.2pc}}
|>{\hspace{0.2pc}}r<{\hspace{0.2pc}}
|}
\hline
\textbf{Age Range} & \textbf{Male} & \textbf{Female}\\
\hline
\hline
under 20 & 52 & 86\\
20-29 & 244 & 210\\
30-39 & 148 & 148\\
40-49 & 148 & 148 \\
50-59 & 148 & 148 \\
60-69 & 171 & 236\\
over 70 & 123 & 57\\
\hline
\end{tabular} \caption{Age Ranges and Gender}
\label{tab:dimFFT}
\end{subtable}%
\hspace*{3em}
\begin{subtable}{0.3\linewidth}
\centering
\begin{tabular}{|c|c|c|}
\hline
\multirow{3}{*}{\textbf{Gender}} & Male & 1034 (50.02 \%)\\ 
& Female  & 1033 (49.98 \%) \\ 
& Actual Male \% & 50 \%~\cite{internet_users_japan}\\
\hline
\multirow{6}{*}{\textbf{Age}} & Average & 44.81 \\ 
& Median & 45 \\ 
& Min & 15 \\ 
& Max & 70 \\ 
& Actual Median & 35 to 44~\cite{internet_users_japan} \\ 
\hline
\end{tabular}
\caption{Matching Actual Statistics}
\label{tab:dimGMM}
\end{subtable}
\end{table*}

\begin{figure}[!ht]
  \label{fig1}
    \centering
    \includegraphics[width=0.55\columnwidth]{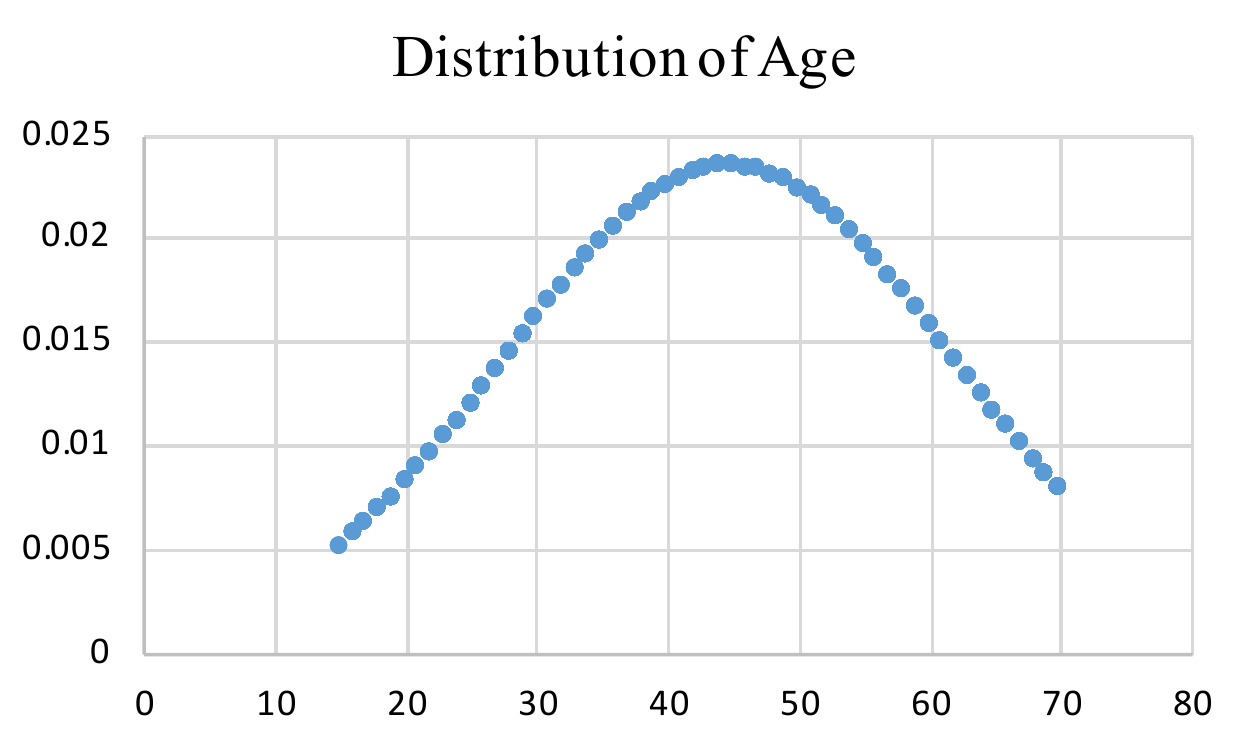}    
    \qquad
    \begin{tabular}[b]{l |r}\hline
    \textbf{Metric} & \textbf{Value}\\
       \hline
      \hline
      Mean & 44.812 \\
      Standard Error & 0.375\\
      Median&  45.000\\
      Standard Deviation &  17.072\\
      Sample Variance & 291.439\\
      Kurtosis & -1.361\\
      Skewness & 0.005 \\ 
      Range & 55.000\\
      Confidence Level (95.0 \%) & 0.736\\
      Count & 2067.000\\
      \hline
    \end{tabular}
    \caption{Distribution Curve and Distribution Summary of The Age}
    \label{fig:table:DistributionAge}
  \end{figure}

As mentioned in Section~\ref{section:procedure}, the 2067 participants are chosen from Internet users in Japan. We match them with a report of the population census from Japanese Internet users~\cite{internet_users_japan}. Table~\ref{table:population} describes the age and gender of our samples. Table~\ref{tab:dimFFT} describes the distribution of gender with different age ranges. Table~\ref{tab:dimGMM} shows the actual percentage of men within the population of Internet users, and the range in which the actual median age of Internet users lies. The normal distribution test is given in Figure~\ref{fig:table:DistributionAge}. The bell curve and the skewness (0.005) that is very close to 0 show that the data is valid for a normal distribution. 

\subsection{Cronbach's Alpha ($\alpha$) Measurement}
We use the Cronbach's $\alpha$~\cite{cronbach1, cronbach2} to measure the internal consistency (IC) or the reliability of the questions that have multiple Likert-scale sub-questions. Suppose that we measure a quantity which is a sum of $K$ components: $X = Y_{1} + Y_{2} + \cdots + Y_{K}$. The Cronbach's $\alpha$ is defined as follows:
\begin{equation}
\alpha = \frac{K}{K-1} (1- \frac{\sum^{K}_{i=1} \sigma^{2}_{Y_{i}}}{\sigma^{2}_{X}}),
\end{equation}
where $\sigma^{2}_{X}$ denotes the variance of the observed total test scores, and $\sigma^{2}_{Y_{i}}$ denotes the variance of the component $i$ for the current sample of persons. We then use the rule of thumb for interpreting $\alpha$ as follows:
\begin{itemize}
\item $\alpha \geq 0.9$: \emph{Excellent} IC
\item $0.9 > \alpha \geq 0.8$: \emph{Good} IC
\item $0.8 > \alpha \geq 0.7$: \emph{Acceptable} IC
\item $0.7 > \alpha \geq 0.6$: \emph{Questionable} IC
\item $0.6 > \alpha \geq 0.5$: \emph{Poor} IC
\item $0.5 > \alpha$: \emph{Unacceptable} IC
\end{itemize}

In our survey, five questions consist of multiple sub-questions. Three of them include brand familiarity (4-point Likert-scale), security behavior (5-point Likert-scale), and security self-confidence (5-point Likert-scale). For the security knowledge (that contains eighteen binary sub-questions) and the user decision on distinguishing eighteen domains (that contains also eighteen binary sub-questions), we consider them as 2-point Liker-Scale questions. The result of Cronbach's $\alpha$ is showed in Table~\ref{table:cronbach}. The internal consistency of all the questions is better than or equal to \emph{acceptable}. This indicates that our survey is reliable.

\begin{table*}
\center
\caption{Cronbach's $\alpha$ Results for Likert-scale Questions}
\begin{tabular}{
|l
|r
|r
|r
|r
|c
|}
\hline
\multirow{ 2}{*}{\textbf{Question}}  & \textbf{No. of Sub-}  & \textbf{Sum of Item Vari-} & \textbf{Variance of Total} & \textbf{Cronbach's}   & \multirow{ 2}{*}{\textbf{IC}} \\
& \textbf{questions} ($K$) &  \textbf{ances} ($\sum^{K}_{i=1} \sigma^{2}_{Y_{i}}$) & \textbf{Scores} ($\sigma^{2}_{X}$) & $\alpha$ &\\
\hline
\hline
Brand Familiarity & {9\ } & {4.446\ } & {12.167\ } & {0.713\ } & Acceptable\\
\hline
Security Behavior & {16\ } & {27.203\ } & {163.219\ } & {0.889\ } & Good \\
\hline
Security Confidence & {6\ } & {5.699\ } & {25.920\ } & {0.936\ } & Excellent\\
\hline
Security Knowledge & {18\ } & {3.038\ } & {10.109\ } & {0.741\ } & Acceptable\\
\hline
Homograph Decision & {18\ } & {2.585\ } & {16.095\ } & {0.889\ } & Good\\
\hline
\end{tabular}
\label{table:cronbach}
\end{table*}

\subsection{Result for Group 1}
When distributing the survey to the participants, we hypothezied that nobody can distinguish the homographs because the visual similarity is almost 100 \%. However, the actual data surprisingly contains a large number of correct answers (over 30 \% for domain \#3 and \#4, and even over 60 \% for domain \#10 and \#15). Fortunately, the analysis of lucky and neutral answers given in Table~\ref{table:lucky} indicates that there is no correct answer with appropriate reasons in the case of domains \#3, \#4 and \#15 which have 100 \% of SSIM, and only 0.19 \% of correct answers with appropriate reasons in the case of domains \#10 which has 99.9 \% of SSIM. We now can confirm that there is no way for the participants to distinguish such extremely high-SSIM homographs. This raises the seriousness of homograph attacks. For this group, we only did the statistics without the need to apply the regression model. 

\subsection{Result for Group 2}
In the first experiment plan (integration), each domain in this group has a different set of incorrect answers and correct answers with appropriate reasons. Finally, 146 common samples (out of 2067 samples) are filtered. The regression model with the target function $f_1$ given in Equation~\ref{eq:target1} is applied and the result is showed in Table~\ref{table:group2}. Remind that, (*) represents $0.01 < p\leq0.05$, (**) represents $0.001 < p\leq0.01$, and (***) represents $p\leq 0.001$. There are four affecting factors found including:
\begin{itemize}
\item Have a job: \emph{normal factor}, the positive coefficient ($0.1425$) indicates that people who have a job tend to have the ability of homograph recognition.
\item Know only Japanese: \emph{semi-significant factor},  the negative coefficient ($-0.2636$) indicates that people who do not only know the Japanese have the ability of homograph recognition.
\item Frequently use the brands: \emph{semi-significant factor}, the positive coefficient ($0.0322$) indicates that people who are more familiar with the brands have the ability of homograph recognition.
\item Have better security knowledge: \emph{significant factor}, the positive coefficient ($0.0624$) indicates that people who have better security knowledge have the ability of homograph recognition.
\end{itemize}

For the second experiment plan (separation), the regression model with the target function $f_2$ given in Equation~\ref{eq:target2} is applied on seven different sets of the incorrect answers and correct answers for appropriate reasons of the seven domains in this group. The factors affecting $f_2$ are found for each domain. The common factors are then extracted. The number of samples in each domain is respectively 965 (\#1), 816 (\#6), 475 (\#7), 650 (\#9), 584 (\#12), 525 (\#14), and 700 (\#18). The result is shown in the last seven columns of Table~\ref{table:group2}. In this table, only the $p$-values of the affecting factors are described so that the common factors can be easily observed. $(+)$ represents the positive coefficients. $(-)$ represents the negative coefficients. The factors chosen for this plan is the common factors that affect more than or equal to $\lceil N/2 \rceil = 4$ domains including:
\begin{itemize}
\item Sex (male): affecting 6/7 domains, is a \emph{significant factor} of 5 domains (\#1, \#9, \#12, \#14, \#17) and a \emph{normal factor} of \#7. All the coefficients are negative, this indicates that the females tend to recognize homographs better than the males. 
\item Have a job: affecting 5/7 domains, is a \emph{significant factor} of \#9, \#17 and a \emph{normal factor} of \#7, \#12 and \#14. All the coefficients are positive; this indicates that the people who have a job tend to be able to recognize the homographs.
\item Still browsing the website even if there is a warning from an anti-virus software: affecting 4/7 domains, is a \emph{semi-significant factor} of \#1 and a \emph{normal factor} of \#12, \#14, and \#17. All the coefficients are negative; this indicates that people who do not browse the website when there is a warning tend to be able to distinguish the homographs.
\item Have more security knowledge: affecting 7/7 domains, is a \emph{significant factor} and has positive coefficients for all the domains. This indicates that the people who have better security knowledge tend to be able to distinguish the homographs.
\end{itemize}

\begin{table*}
\center
\caption{Experiment Result of Group 2 (Homograph with SSIM $<0.999$)}
\begin{tabular}{
|c|l||
r|r|r||
c|c|c|c|c|c|c|}
\hline
\multirow{ 2}{*}{\textbf{No.}} & \multirow{ 2}{*}{\textbf{Factors}} & \multicolumn{3}{c||}{\textbf{Integration}} & \multicolumn{7}{c|}{\textbf{Separation}}\\
\cline{3-12}
& & \textbf{Coef.} & \textbf{$p$} & \textbf{95\%CI} & \textbf{\#1} & \textbf{\#6} & \textbf{\#7} & \textbf{\#9} & \textbf{\#12} & \textbf{\#14} & \textbf{\#17}\\
\hline
\hline
\multicolumn{2}{|c||}{Number of Samples} &  \multicolumn{3}{|c||}{146}& 965 & 816& 475& 650& 584& 525& 700\\
\hline
\multicolumn{2}{|c||}{(Intercept)} &  -0.6607 & 0.007   & [-1.134, -0.187] &  & & & & && \\\hline
1& Sex (male) &  -0.0705 &  0.261 & [-0.194, \hphantom{-}0.053] & \cellcolor{lightgray}{\textless 0.001} &  \cellcolor{lightgray}{}  & \cellcolor{lightgray}{0.032} & \cellcolor{lightgray}{0.001} & \cellcolor{lightgray}{\textless 0.001} & \cellcolor{lightgray}{\textless 0.001} &  \cellcolor{lightgray}{0.001} \\
& & &  &  & \cellcolor{lightgray}{$(-)$***}  & \cellcolor{lightgray}{} &  \cellcolor{lightgray}{$(-)$*} & \cellcolor{lightgray}{$(-)$***} & \cellcolor{lightgray}{$(-)$***} & \cellcolor{lightgray}{$(-)$***} & \cellcolor{lightgray}{$(-)$***} \\
\hline
2& Age (older) & -0.0022 &   0.251 & [-0.006, \hphantom{-}0.002] & & &  &   &  \textless 0.001&  0.001 &  \textless 0.001\\
& & &  &  & &  & & & $(-)$*** & $(-)$***& $(-)$***\\
\hline
3& Have a job&  \cellcolor{lightgray}{0.1425} &  \cellcolor{lightgray}{0.036} & \cellcolor{lightgray}{[\hphantom{-}0.009, \hphantom{-}0.276]} & \cellcolor{lightgray}{} & \cellcolor{lightgray}{}&   \cellcolor{lightgray}{0.016}  &  \cellcolor{lightgray}{\textless 0.001}&  \cellcolor{lightgray}{0.015}  &  \cellcolor{lightgray}{0.043}&  \cellcolor{lightgray}{\textless 0.001}\\
 & & \cellcolor{lightgray}{} &  \cellcolor{lightgray}{*} & \cellcolor{lightgray}{} & \cellcolor{lightgray}{}& \cellcolor{lightgray}{}& \cellcolor{lightgray}{$(+)$*} & \cellcolor{lightgray}{$(+)$***} &  \cellcolor{lightgray}{$(+)$*} & \cellcolor{lightgray}{$(+)$*} & \cellcolor{lightgray}{$(+)$***}\\
\hline
4& Know only Japanese &  -0.2636 & 0.006 & [-0.451, -0.077] & & &   &  & 0.015 & \textless 0.001 &  0.007\\
& & & ** &  &  & && & $(-)$*& $(-)$***& $(-)$**\\
\hline
5& Number of languages &  0.0262 & 0.519  & [-0.054, \hphantom{-}0.106] & & & & & &&\\
\hline
6& Install anti-virus &  -0.0920 & 0.189 & [-0.230, \hphantom{-}0.046] & & & & & && \\
\hline
7&Browse even warning & -0.0295 &  0.648 & [-0.157, \hphantom{-}0.098] &  0.010 &  & &   & 0.018 &  0.042 &  0.044\\
& & &  &  & $(-)$** & & &  & $(-)$* & $(-)$*& $(-)$*\\
\hline
8& Frequently use brands & 0.0322 & 0.004 & [\hphantom{-}0.010, \hphantom{-}0.054] &  0.001 & & 0.010& &   & &  \\
& & & ** &  &  $(+)$***&  & $(-)$** & &  & & \\
\hline
9& Education in CS/CE & -0.0316 & 0.820 & [-0.306, \hphantom{-}0.243] &  & & & & &&\\
\hline
10& Work in CS/CE & -0.1940 &  0.093  & [-0.421, \hphantom{-}0.033] &   & & & & &    & \\
\hline
11& More sec. behavior &  0.0009 & 0.753 &  [-0.005, \hphantom{-}0.007] & 0.043& & & & &&\\
& & &  &  & $(+)$* & & & & &&\\
\hline
12 & More sec. knowledge & \cellcolor{lightgray}{0.0624} &  \cellcolor{lightgray}{\textless 0.001} & \cellcolor{lightgray}{[\hphantom{-}0.041, \hphantom{-}0.084]} &  \cellcolor{lightgray}{\textless 0.001} &   \cellcolor{lightgray}{\textless 0.001} & \cellcolor{lightgray}{\textless 0.001}  &  \cellcolor{lightgray}{\textless 0.001}  &  \cellcolor{lightgray}{\textless 0.001} &  \cellcolor{lightgray}{\textless 0.001} & \cellcolor{lightgray}{\textless 0.001}\\
& & \cellcolor{lightgray}{}& \cellcolor{lightgray}{***} & \cellcolor{lightgray}{}& \cellcolor{lightgray}{$(+)$***} & \cellcolor{lightgray}{$(+)$***}  & \cellcolor{lightgray}{$(+)$***}  & \cellcolor{lightgray}{$(+)$***}  & \cellcolor{lightgray}{$(+)$***} &  \cellcolor{lightgray}{$(+)$***} &  \cellcolor{lightgray}{$(+)$***}\\
\hline
13& More sec. confidence &  -0.0007 &  0.930  & [-0.015, \hphantom{-}0.014] & & &   & & &&  \\
\hline
\end{tabular}
`*': $0.01 < p \leq 0.05$, `**': $0.001 < p \leq 0.01$, `***': $p \leq 0.001$ \\
$(+)$: coefficient $>0$, $(-)$: coefficient $<0$
\label{table:group2}
\end{table*}

\subsubsection{Consistency}
The results of the plans are not the same, so we perform the result consistency as indicated in Section~\ref{section:consistency}. The final set we are aiming to find (filled with gray color in Table~\ref{table:group2}) consists of the following affecting factors:
\begin{itemize}
\item Sex (female) since all the coefficients are negative for males in the separation plan.
\item People who have a job: this is the common factor of both the plans.
\item More security knowledge: this is also the common factor of both the plans.
\end{itemize}

\subsection{Result for Group 3}
In this group, as explained in Section~\ref{section:luckyneutral}, the lucky and neutral answers are not necessary to be excluded from the dataset. For the both experiment plans (integration and separation), the regression model is applied on all 2067 samples using the target function $f_1$ (Equation~\ref{eq:target1}) in the case of integration and $f_2$ (Equation~\ref{eq:target2}) in the case of separation. The results are showed in Table~\ref{table:group3}.

For the first experiment plan (integration), there are seven affecting factors found including:
\begin{itemize}
\item Sex (male): \emph{normal factor}, the positive coefficient ($ 0.1354$) indicates that the males tend to be able to distinguish the non-homographs.
\item Age (older): \emph{semi-significant factor}, the negative factor ($-0.0052$) indicates that the young people tend to be able to distinguish the non-homographs.
\item Have a job: \emph{significant factor}, the negative factor ($-0.2104$) indicates that the people who do not have a job tend to be able to distinguish the non-homographs.
\item  Browsing the website even if there is a warning from an anti-virus software: \emph{semi-significant factor}, the positive coefficient ($0.1484>0$) indicates that the people who still browse the website even if there is a warning tend to be able to distinguish the non-homographs.
\item Education in CS/CE: \emph{normal factor}, the positive coefficient ($0.3072$) indicates that the people who are educated in computer science or computer engineering tend to be able to distinguish the non-homographs. 
\item Work in CS/CE: \emph{normal factor}, the positive coefficient ($0.2861$) indicates that the people who work in computer science or computer engineering tend to be able to distinguish the non-homographs. 
\item Have better security knowledge: \emph{significant factor}, the negative coefficient ($-0.0551$) indicates that people who have less security knowledge have better ability in distinguishing the non-homographs.
\end{itemize}

For the second experiment plan (separation), since this group also has seven domains as the group 2, the factors chosen for this plan are the common factors that affect more than or equal to 4 domains.
\begin{itemize}
\item Sex (male): affecting 4/7 domains, \emph{significant factor} of \#2, \emph{semi-significant factor} of \#8, \emph{normal factor} of \#5 and \#18. All the coefficients are positive; this indicates that the males tend to be able to distinguish the non-homographs.
\item Age (older): affecting 4/7 domains, \emph{significant factor} of \#11 and \#13, \emph{semi-significant factor} of \#8 and \#16. All the coefficients are negative; this indicates that the young people tend to be able to distinguish the non-homographs.
\item Have a job: affecting 6/7 domains, \emph{semi-significant factor} of \#8, \#13, and \#16, \emph{normal factor} of \#2, \#11, and \#18. All the coefficients are negative; this indicates that the people who do not have a job tend to be able to distinguish the non-homographs.
\item Have better security knowledge: affecting 5/7 domains, is a \emph{significant factor} and has negative coefficients for all the domains. This indicates that the people who have less security knowledge tend to be able to distinguish the non-homographs.
\end{itemize}

\begin{table*}[!ht]
\center
\caption{Experiment Result of Group 3 (Non-homograph)}
\begin{tabular}{|
c|l|
r|r|r||
c|c|c|c|c|c|c
|}
\hline
\multirow{ 2}{*}{\textbf{No.}} & \multirow{ 2}{*}{\textbf{Factors}} & \multicolumn{3}{c||}{\textbf{Integration}} & \multicolumn{7}{c|}{\textbf{Separation}}\\
\cline{3-12}
& & \textbf{Coef.} & \textbf{$p$} & \textbf{95\%CI} & \textbf{\#2} & \textbf{\#5} & \textbf{\#8} & \textbf{\#11} & \textbf{\#13} & \textbf{\#16} & \textbf{\#18}\\
\hline
\hline
\multicolumn{2}{|c|}{No. of Samples} &  \multicolumn{10}{|c|}{2067}\\
\hline
\multicolumn{2}{|c|}{(Intercept)}& 1.3626 &   \textless 0.001 & [\hphantom{-}0.934, \hphantom{-}1.791] & & & & & && \\
\hline
1& Sex (male) & \cellcolor{lightgray}{0.1354} & \cellcolor{lightgray}{0.012} & \cellcolor{lightgray}{[\hphantom{-}0.029, \hphantom{-}0.241]} & \cellcolor{lightgray}{\textless 0.001} &  \cellcolor{lightgray}{0.019} & \cellcolor{lightgray}{0.004} & \cellcolor{lightgray}{} & \cellcolor{lightgray}{} & \cellcolor{lightgray}{} & \cellcolor{lightgray}{0.011}\\
& & \cellcolor{lightgray}{} & \cellcolor{lightgray}{*} & \cellcolor{lightgray}{} & \cellcolor{lightgray}{$(+)$***} & \cellcolor{lightgray}{$(+)$*} & \cellcolor{lightgray}{$(+)$**} & \cellcolor{lightgray}{} & \cellcolor{lightgray}{} & \cellcolor{lightgray}{} & \cellcolor{lightgray}{$(+)$*}\\
\hline
2& Age (older) & \cellcolor{lightgray}{-0.0052} &  \cellcolor{lightgray}{0.002} & \cellcolor{lightgray}{[-0.008, -0.002]} & \cellcolor{lightgray}{} & \cellcolor{lightgray}{} &  \cellcolor{lightgray}{0.002} &   \cellcolor{lightgray}{\textless 0.001} & \cellcolor{lightgray}{\textless 0.001} &  \cellcolor{lightgray}{0.003} & \cellcolor{lightgray}{}\\
& & \cellcolor{lightgray}{} &  \cellcolor{lightgray}{**} & \cellcolor{lightgray}{}  & \cellcolor{lightgray}{} & \cellcolor{lightgray}{} & \cellcolor{lightgray}{$(-)$**} & \cellcolor{lightgray}{$(-)$***} & \cellcolor{lightgray}{$(-)$***} & \cellcolor{lightgray}{$(-)$**} & \cellcolor{lightgray}{}\\
\hline
3& Have a job& \cellcolor{lightgray}{-0.2104} &  \cellcolor{lightgray}{\textless 0.001} & \cellcolor{lightgray}{[-0.326, -0.095]} & \cellcolor{lightgray}{0.011} & \cellcolor{lightgray}{} &  \cellcolor{lightgray}{0.007} &  \cellcolor{lightgray}{0.020} & \cellcolor{lightgray}{0.009} & \cellcolor{lightgray}{0.007} & \cellcolor{lightgray}{0.027}\\
& & \cellcolor{lightgray}{} & \cellcolor{lightgray}{***} & \cellcolor{lightgray}{} & \cellcolor{lightgray}{$(-)$*} & \cellcolor{lightgray}{} & \cellcolor{lightgray}{$(-)$**} & \cellcolor{lightgray}{$(-)$*} & \cellcolor{lightgray}{$(-)$**} & \cellcolor{lightgray}{$(-)$**} & \cellcolor{lightgray}{$(-)$*}\\
\hline
4& Know only  & 0.0931 & 0.258 & [-0.068, \hphantom{-}0.254] & & &   & 0.014 & &&\\
& Japanese & &  &  &  & && $(+)$*& &&\\
\hline
5& Number of  & 0.0087 & 0.830 & [-0.071, \hphantom{-}0.088] & & & & & &&\\
& languages & &  &  &  & && & &&\\
\hline
6& Install & -0.0796 & 0.211 & [-0.204, \hphantom{-}0.045] & & & & & && 0.043\\
& anti-virus & &  &  &  & & & & && $(-)$*\\
\hline
7&Browse even & 0.1484 & 0.009 & [\hphantom{-}0.037, \hphantom{-}0.259] & & 0.001 & &  & 0.040 &&\\
& warning & &  **&  & & $(+)$***& & & $(+)$*&&\\
\hline
8& Frequently & 0.0149 & 0.138 & [-0.005, \hphantom{-}0.035] & & & & &  \textless 0.001 &&\\
& use brands & &  &  &  & & & & $(+)$*** &&\\
\hline
9& Education & 0.3072 & 0.035 & [\hphantom{-}0.021, \hphantom{-}0.593] & 0.016 & & & & &&\\
& in CS/CE & & * &  & $(+)$* & & & & &&\\
\hline
10& Work & 0.2861 &  0.018 & [\hphantom{-}0.049, \hphantom{-}0.523] &  0.002 & & & & && 0.010\\
& in CS/CE & & * &  & $(+)$** & & & & && $(+)$**\\
\hline
11& More sec.  &  -0.0020 & 0.412 &  [-0.007, \hphantom{-}0.003] & & & & & &&\\
& behavior & &  &  &  & && & &&\\
\hline
12 & More sec.  & \cellcolor{lightgray}{-0.0551} & \cellcolor{lightgray}{\textless 0.001} & \cellcolor{lightgray}{[-0.075, -0.035]} & \cellcolor{lightgray}{\textless 0.001} & \cellcolor{lightgray}{\textless 0.001} & \cellcolor{lightgray}{\textless 0.001} &  \cellcolor{lightgray}{0.001} & \cellcolor{lightgray}{} & \cellcolor{lightgray}{} & \cellcolor{lightgray}{\textless 0.001}\\
& knowledge & \cellcolor{lightgray}{}  & \cellcolor{lightgray}{} {***} &  \cellcolor{lightgray}{}{} & \cellcolor{lightgray}{} {$(-)$***} & \cellcolor{lightgray}{} {$(-)$***} & \cellcolor{lightgray}{} {$(-)$***} & \cellcolor{lightgray}{} {$(-)$***} & \cellcolor{lightgray}{} {} & \cellcolor{lightgray}{} {} & \cellcolor{lightgray}{} {$(-)$***} \\
\hline
13& More sec.  & 0.0060 &  0.322 & [-0.006, \hphantom{-}0.018] & & &   & & &&  0.049\\
& confidence & &  &  & & & & & && $(+)$*\\
\hline
\end{tabular}
`*': $0.01 < p \leq 0.05$, `**': $0.001 < p \leq 0.01$, `***': $p \leq 0.001$\\
$(+)$: coefficient $>0$, $(-)$: coefficient $<0$
\label{table:group3}
\end{table*}

\subsubsection{Consistency}
Similar to the group 2, the results of the plans in this group are also not the same, so we perform the result consistency as indicated in Section~\ref{section:consistency}. The final set we are aiming to find (filled with the gray color in Table~\ref{table:group3}) consists of the following affecting factors. Fortunately, all the factors are the common factors of both the plans.
\begin{itemize}
\item Sex (male): since all the coefficients are positive for both the plans.
\item Young people: since all the coefficients are negative.
\item People who do not have a job: since all the coefficients are negative.
\item Less security knowledge: since all the coefficients are negative.
\end{itemize}

\section{Discussion}
\label{section:discussion}
In this section, we discuss how the factors change when the participants are explained about the homographs. We then discuss some several ideas for improving the result and their challenges for future work.

\subsection{Before and after Homograph Explanation/Education}
The main result is described in Section~\ref{section:experiment}. In this section, we perform an extra analysis of how the factors change when the participants are explained about what the homograph attack is. In the survey, after the participants give their decisions to the eighteen domains, a description of the homograph is displayed (Appendix~\ref{section:explain}). The participants then respond to their decision again to the same eighteen domains. To avoid data outlier in the participants' decisions (for ensuring the independency in their decision before and after the homograph explanation), the web interface of the survey is designed so that the participants cannot go back to previous questions before the homograph explanation from the questions that are displayed after the homograph explanation. 

In this analysis, we consider the integration plan for all the eighteen domains with all the 2067 participants\footnote{Although there are lucky and neutral answers, they actually happened (these answers are the actual samples in the dataset) and we would want to know how the factors are in this extra analysis.}. Table~\ref{table:beforeafter} shows the experiment result. $p_{BE}$ and $p_{AF}$ denotes the $p$-values before and after the homograph explanation, respectively. $\mid \bigtriangleup p\mid$ denotes the change's magnitude of $p$. The fifth and sixth columns are the change of the coefficient signs and the significane, respectively. N/A in the sixth column means that the factor is not an affecting factor (e.g., ** $\rightarrow$ N/A means the variable is a semi-significant factor before the homograph explanation, but after that, it is no longer an affecting factor). The result shows that there are five factors found in both cases of before and after the homograph explanation. The three factors (anti-virus installation, frequently use brands and more security knowledge) are consistent for both the cases. Sex (male) is no longer an affecting factor after the homograph explanation. Interestingly, working in computer science or computer engineering from not an affecting factor becomes an affecting factor. Furthermore, $\mid \bigtriangleup p\mid= 0.061$ is highest compared to other affecting factors. This indicates that people who work in computer science or computer engineering are able to capture the situation quickly after being explained about the homographs.

\begin{table*}
\center
\caption{Factors Change before and after the Homograph Explanation}
\begin{tabular}{
|>{\hspace{0.2pc}}l<{\hspace{0.2pc}}
|>{\hspace{0.2pc}}r<{\hspace{0.2pc}}
|>{\hspace{0.2pc}}r<{\hspace{0.2pc}}
|>{\hspace{0.2pc}}r<{\hspace{0.2pc}}
|>{\hspace{0.2pc}}c<{\hspace{0.2pc}}
|>{\hspace{0.2pc}}c<{\hspace{0.2pc}}
|}
\hline
\textbf{Factors} & $p_{BE}$ & $p_{AF}$ & $\mid \bigtriangleup p \mid$ & \textbf{Coefficient Sign}& \textbf{Significancy}   \\
\hline
\hline
Sex (male) &  0.018 &  0.339 & 0.321 & $(+) \rightarrow (-)$ & ** $\rightarrow$ N/A \\
\hline
Install anti-virus & 0.001 & 0.045  & 0.044 & $(-) \rightarrow (-)$ & *** $\rightarrow$  *\\
\hline
Frequently use brands & \textless 0.001 & \textless 0.001 & \textless 0.001 & $(-) \rightarrow (-)$ &  *** $\rightarrow$  ***\\
\hline
Work in CS/CE & 0.083 &  0.022  & 0.061 & $(+) \rightarrow (+)$ & N/A $\rightarrow$ * \\
\hline
More security knowledge &  \textless 0.001 & \textless 0.001 & \textless 0.001 & $(-) \rightarrow (-)$ & *** $\rightarrow$ *** \\
\hline
\end{tabular}
\label{table:beforeafter}
\end{table*}

\subsection{Future Work and Challenges}
Related to the survey itself, there are three ideas that can improve the study. First, in this current work, the survey is applied for local participants (i.e., Japanese). If it can be applied for global participants, the responses would be more objective. In this case, there is a challenge in translating the survey across the languages in different countries. The translation should be appropriately considered while preserving its reliability and structure validity. Second, some features which may affect the ability of homograph recognition including how many hours for using the Internet per day, factors related to participant psychology like emotional state, demands and the environment when answering the questionnaire, etc. Third, if the domains are asked to the participants in an actual simulation rather than in a self-report questionnaire, the bias can be reduced and also other information related to participants can be extracted such as the time of accessing domains, the scenario of accessing domains, and the mouse move.

Related to the model, some promising elements can be included in the target functions. The first is the Alexa ranking. Some domains are very famous (e.g., \url{amazon.com} or \url{google.com}), and thus the participants are more familiar with them rather than the domains that are less popular (e.g., \url{coinbase.com}). The Alexa ranking can be considered in a global scope (if the survey is applied to different countries) or in a local scope (if the survey is applied to a country like this work). The second is the order of the domains in the questionnaire. In fact, the participants tend to carefully answer the first few domains but gradually tend to answers the domains randomly; and therefore there is bias in this case. The domain order in the questions should be added as a component in the target function. Furthermore, there can be another bias when the participants answer all domains as homographs because they perhaps think that it has a high probability for the domains to be homographs in such a security survey, or think that false positive is better than false negative when they are not sure. Designing a survey that can eliminate data bias is a challenge in most of human factor research topics.


\section{Conclusions}
\label{section:conclusion}
We designed and ran an online study to explore how user demographics, brand familiarity, and security backgrounds affect the ability in recognizing homographs. We collected 2,067 responses to our survey from participants located in Japan and analyzed them using linear regression. Our results shed light on the differences in the ability of homograph recognition for different kinds of homographs. We find that 13.95\% of participants can recognize non-homographs while 16.60\% of participants can recognize homographs when the visual similarity with the target brand domains is under 99.9\%; but when the similarity increases to 99.9\%, the number of participants who can recognize homographs significantly drops down to only 0.19\%; and for the homographs with 100\% of visual similarity, there is no way for the participants to recognize. We also find that for different levels of visual similarity, the participants exhibit different abilities. Female participants tend to recognize homographs while male participants tend to able to recognize non-homographs. Security knowledge is a significant factor affecting both homographs and non-homographs. Surprisingly, people who have strong security knowledge tend to be able to recognize homographs but not non-homographs. Furthermore, an interesting result is that people who work or are educated in computer science or computer engineering is not an affecting factor for the ability in recognizing homograph as hypothesized; however, right after being explained about homograph attack, they are the ones who can capture the situation the most quickly. 

For the implication, first, we want to raise the seriousness of the homograph attack. Second, we want to recommend looking into directions beyond user education to promote more ability in homograph recognition, especially aiming at people who are male, who do not have a job, and who have less security knowledge. Third, we want to emphasize that not all the domains that have high visual similarity with the brand domains are the homographs. User education for non-homographs is also necessary and can be aimed at people especially those who are female, elder, have a job and have good security knowledge.

\appendix

\section{Appendix: Security Behavior} 
\label{section:security_behavior}
The question of security behavior consists of the following sixteen sub-questions:
\begin{enumerate}
\item I set my computer screen to automatically lock if I don't use it for a prolonged period of time.
\item I use a password/passcode to unlock my laptop or tablet.
\item I manually lock my computer screen when I step away
from it.
\item I use a PIN or passcode to unlock my mobile phone.
\item I change my passwords even if it is not needed.
\item I use different passwords for different accounts that I
have.
\item When I create a new online account, I try to use a password that goes beyond the site's minimum.
\item I include special characters in my password even if it's not required.
requirements.
\item When someone sends me a link, I open it only after verifying where it goes.
\item I know what website I'm visiting by looking at the URL bar, rather than by the website's look and feel.
\item I verify that information will be sent securely (e.g., SSL, ``https://", a lock icon) before I submit it to websites.
\item When browsing websites, I mouseover links to see where they go, before clicking them.
\item If I discover a security problem, I fix or report it rather than assuming somebody else will.
\item When I'm prompted about a software update, I install it right away.
\item I try to make sure that the programs I use are up-to-date.
\item I verify that my anti-virus software has been regularly updating itself.
\end{enumerate}

\subsubsection{Answer Options} There are five answer options for each sub-question. The order numbers are also the actual values used in the experiment.
\begin{enumerate}
\item Not at all
\item Not much
\item Sometimes
\item Often
\item Always
\end{enumerate}

\section{Appendix: Security Knowledge} 
\label{section:security_knowledge}
The question of security knowledge consists of the following eighteen sub-questions:
\begin{enumerate}
\item My Internet provider and location can be disclosed from my IP address.
\item My telephone number can be disclosed from my IP addresses.
\item The web browser information of my device can be disclosed to the operators of websites.
\item Since Wi-Fi networks in coffee shops are secured by the coffee shop owners, I can use them to send sensitive data such as credit card information.
\item Password comprised of random characters are harder for attackers to guess than passwords comprised of common words and phrases.
\item If I receive an email that tells me to change my password, and links me to the web page, I should change my password immediately.
\item My devices are safe from being infected while browsing the web because web browsers only display information.
\item It is impossible to confirm whether secure communication is being used between my device and a website.
\item My information can be stolen if a website that I visit masquerades as a famous website (e.g., amazon.com).
\item I may suffer from monetary loss if a website that I visit masquerades as a famous website.
\item My devices and accounts may be put at risk if I make a typing mistake while entering the address of a website.
\item My IP address is secret and it is unsafe to share it with anyone.
\item If my web browser does not show a green lock when I visit a website, then I can deduce that the website it is malicious.
\item It is safe to open links that appear in emails in my inbox.
\item It is safe to open attachments received via email.
\item I use private browsing mode to protect my machine from being infected.
\item It is safe to use anti-virus software downloaded through P2P file sharing services.
\item Machines are safe from infections unless participants actively download malware.
\end{enumerate}

\subsubsection{Answer Options} There are two answer options for each sub-question. The order numbers are also the actual values used in the experiment.
\begin{enumerate}
\item True (the value used in the experiment: 1)
\item False (the value used in the experiment: 0)
\end{enumerate}

\subsubsection{Correct Answers}
The correct answers for the eighteen sub-questions are: \emph{true} for sub-questions 1, 3, 5, 9, 10, 11, and \emph{false} for the others.

\section{Appendix: Security Self-Confidence}
\label{section:security_confidence}
The question of security self-confidence consists of the following six sub-questions:
\begin{enumerate}
\item I know about countermeasures for keeping the data on my device from being exploited.
\item I know about countermeasures to protect myself from monetary loss when using the Internet.
\item I know about countermeasures to prevent my IDs or Passwords being stolen.
\item I know about countermeasures to prevent my devices from being compromised.
\item I know about countermeasures to protect me from being deceived by fake web sites.
\item I know about countermeasures to prevent my data from being stolen during web browsing.
\end{enumerate}

\subsubsection{Answer Options}
There are five answer options for each sub-question. The order numbers are also the actual values used in the experiment.
\begin{enumerate}
\item Not at all
\item Not applicable
\item Neither agree nor disagree
\item Applicable
\item Very applicable
\end{enumerate}

\section{Appendix: Ability of Homograph Recognition}
\label{section:user_decision}
The question of homograph recognition consists of the following eighteen sub-questions:
\begin{enumerate}
\item Domain \#1: \url{xn--mazon-zjc.com} (displayed as the sample 1 in Figure~\ref{fig:user_decision}).
\item Domain \#2: \url{amazonaws.com}.
\item Domain \#3: \url{xn--mazon-3ve.com} (displayed as the sample 3 in Figure~\ref{fig:user_decision}). 
\item Domain \#4: \url{xn--gogle-m29a.com} (displayed as the sample 4 in Figure~\ref{fig:user_decision}). 
\item Domain \#5: \url{google.com.vn}. 
\item Domain \#6: \url{goole.co.jp}. 
\item Domain \#7: \url{xn--coinbas-z8a.com} (displayed as the sample 7 in Figure~\ref{fig:user_decision}).
\item Domain \#8: \url{wikimedia.org}.
\item Domain \#9: \url{xn--wikipdia-f1a.org} (displayed as the sample 9 in Figure~\ref{fig:user_decision}). 
\item Domain \#10: \url{xn--bookin-n0c.com} (displayed as the sample 10 in Figure~\ref{fig:user_decision}). 
\item Domain \#11: \url{jbooking.jp}.
\item Domain \#12: \url{xn--expeda-fwa.com} (displayed as the sample 12 in Figure~\ref{fig:user_decision}). 
\item Domain \#13: \url{expedia.co.jp}.
\item Domain \#14: \url{xn--paypl-6qa.com} (displayed as the sample 14 in Figure~\ref{fig:user_decision}).
\item Domain \#15: \url{xn--pypal-4ve.com} (displayed as the sample 15 in Figure~\ref{fig:user_decision}).
\item Domain \#16: \url{sex.com}.
\item Domain \#17: \url{faeceb0ok.com}.
\item Domain \#18: \url{vi-vn.facebook.com}.
\end{enumerate}

\subsubsection{Answer Questions}
There are two answer options for each sub-question. The order numbers are also the actual values used in the experiment.
\begin{enumerate}
\item Homograph (the value used in the experiment: 1)
\item Non-homograph (the value used in the experiment: 0)
\end{enumerate}

\subsubsection{Correct Answers}
The eighteen domains are displayed respectively in Figure~\ref{fig:user_decision}. The correct answers for the eighteen domains are as follows:
\begin{itemize}
\item Homograph: the domains \#1, \#3, \#4, \#6, \#7, \#9, \#10, \#12, \#14, \#15, \#17.
\item Non-homographs: the others.
\end{itemize}
The homographs \#1 and \#3 target to the brand Amazon. The homographs \#4 and \#6 target to the brand Google. The homograph \#7 targets to the brand Coinbase; the homograph \#9 targets to the brand Wikipedia. The homograph \#10 targets to the brand Booking. The homograph \#12 targets to the brand Expedia. The homographs \#14 and \#15 target to the brand Paypal. The homograph \#17 targets to the brand Facebook.

\section{Appendix: Homograph Explanation}
\label{section:explain}
The description about the homograph attack is given as follows:

``Homograph attack is a way that the attackers deceive victims about what domain they are communicating with by exploiting the fact that many domains look alike. There are several kinds of homographs in the wild, we thus synthesize them into 5 categories. The first is visual homograph which uses different characters but visually look alike, for example: facebook.com and facebôok.com. The second is semantic homograph which use synonyms or contextual similar words, for example: facebook.com and markzuckerbergsocialnetwork.com. The third is TLD homograph which uses the same main domain names, but different the top-level-domain (TLD), for example: facebook.com and facebook.biz. The fourth is typosquatting which relies on mistakes such as typos made by Internet users when typing the domain names, for example: facebook.com and faceboook.com. The last is the combination of the previous 4 categories. Also, note that the homographs in which certain characters are inserted or replaced (known as bitsquatting) in the brand domains are listed in the fourth type (typosquatting homograph); for instance, travelgoogle.com targeting to google.com''.

\end{document}